\documentclass[journal]{IEEEtran}
\usepackage{cite}

\usepackage{graphicx}
\ifCLASSINFOpdf
\else
\fi
\usepackage{amssymb}
\usepackage[cmex10]{amsmath}
\newcounter{MYtempeqncnt}
\usepackage[font=scriptsize]{caption}
\usepackage{color}
\usepackage{fouriernc}
\usepackage{graphicx}
\usepackage{caption}
\usepackage{subcaption}
    
\usepackage{mathtools}

\usepackage[english]{babel}
\usepackage[utf8]{inputenc}
\usepackage{algorithm}

\usepackage[]{algpseudocode}

\algnewcommand\algorithmicinput{\textbf{Input:}}
\algnewcommand\Input{\item[\algorithmicinput]}
\algnewcommand\algorithmicoutput{\textbf{Output:}}
\algnewcommand\Output{\item[\algorithmicoutput]}
\algnewcommand\algorithmichline{}
\algnewcommand\Hline{\item[\algorithmichline]}
\makeatletter
    \newcommand*{\algrule}[1][\algorithmicindent]{\makebox[#1][l]{\hspace*{.5em}\thealgruleextra\vrule height \thealgruleheight depth \thealgruledepth}}%
\newcommand*{\thealgruleextra}{}
\newcommand*{\thealgruleheight}{1.05\baselineskip}
\newcommand*{\thealgruledepth}{.25\baselineskip}

\newcount\ALG@printindent@tempcnta
\def\ALG@printindent{%
    \ifnum \theALG@nested>0
        \ifx\ALG@text\ALG@x@notext
        \else
            \unskip
            \addvspace{-1pt}
            \ALG@printindent@tempcnta=1
            \loop
                \algrule[\csname ALG@ind@\the\ALG@printindent@tempcnta\endcsname]%
                \advance \ALG@printindent@tempcnta 1
            \ifnum \ALG@printindent@tempcnta<\numexpr\theALG@nested+1\relax
            \repeat
        \fi
    \fi
    }%
\usepackage{etoolbox}
\patchcmd{\ALG@doentity}{\noindent\hskip\ALG@tlm}{\ALG@printindent}{}{\errmessage{failed to patch}}
\makeatother

\newbox\statebox
\newcommand{\myState}[1]{%
    \setbox\statebox=\vbox{#1}%
    \edef\thealgruleheight{\dimexpr \the\ht\statebox+1pt\relax}%
    \edef\thealgruledepth{\dimexpr \the\dp\statebox+1pt\relax}%
    \ifdim\thealgruleheight<.75\baselineskip
        \def\thealgruleheight{\dimexpr .75\baselineskip+1pt\relax}%
    \fi
    \ifdim\thealgruledepth<.25\baselineskip
        \def\thealgruledepth{\dimexpr .25\baselineskip+1pt\relax}%
    \fi
    \State #1%
    \def\thealgruleheight{\dimexpr .75\baselineskip+1pt\relax}%
    \def\thealgruledepth{\dimexpr .25\baselineskip+1pt\relax}%
}

\def\NoNumber#1{{\def\alglinenumber##1{}\State #1}\addtocounter{ALG@line}{-1}}

\usepackage{color}

\begin{document}
%
\title{Analysis and Design of Adaptive OCDMA Passive Optical Networks}
%
%
%

\author{Mohammad~Hadi,~\IEEEmembership{Member,~IEEE,}
        and~Mohammad~Reza~Pakravan,~\IEEEmembership{Member,~IEEE}
        
\thanks{Mohammad Hadi is a PhD student at department of Electrical Engineering, Sharif University of Technology, e-mail: mhadi@ee.sharif.edu.}
\thanks{Mohammad Reza Pakravan is faculty member of Electrical Engineering department, Sharif University of Technology, e-mail: pakravan@sharif.edu.}}

\maketitle

\begin{abstract}
\boldmath
OCDMA systems can support multiple classes of service by differentiating code parameters, power level and diversity order. In this paper, we analyze BER performance of a multi-class 1D/2D OCDMA system and propose a new approximation method that can be used to generate accurate estimation of system BER using a simple mathematical form. The proposed approximation provides insight into proper system level analysis, system level design and sensitivity of system performance to the factors such as code parameters, power level and diversity order. Considering code design, code cardinality and system performance constraints, two design problems are defined and their optimal solutions are provided. We then propose an adaptive OCDMA-PON that adaptively shares unused resources of inactive users among active ones to improve upstream system performance. Using the approximated BER expression and defined design problems, two adaptive code allocation algorithms for the adaptive OCDMA-PON are presented and their performances are evaluated by simulation. Simulation results show that the adaptive code allocation algorithms can increase average transmission rate or decrease average optical power consumption of ONUs for dynamic traffic patterns. According to the simulation results, for an adaptive OCDMA-PON with BER value of $10^{-7}$ and user activity probability of $0.5$, transmission rate (optical power consumption) can be increased (decreased) by a factor of $2.25$ ($0.27$) compared to fixed code assignment.
\end{abstract}
%
\IEEEpeerreviewmaketitle
\section{Introduction}\label{sec_I}
\IEEEPARstart{O}{ptical} Code Division Multiple-Access (OCDMA) is a well-known multiple access technique considered in optical networks. In OCDMA networks, users share a common physical channel and each user sends its data using a code string. OCDMA has some advantages over other multi-access methods such as simple and asynchronous operation, high resource utilization and capability of providing differentiated classes of Quality of Service (QoS) for different users. The performance of OCDMA networks is mainly limited by Multi-Access Interference (MAI) from other users \cite{new_ref_9, new_ref_10, new_ref_12, ref_3, ref_pak}.  There are research results that have combined the concepts of Orthogonal Frequency Division Multiple-Access (OFDMA) and OCDMA (known as Multi-Carrier OCDMA) \cite{new_ref_12, new_ref_13} or Wavelength Division Multiplexing and OCDMA (known as WDM/OCDMA) to increase the system performance and capacity \cite{ref_6, ref_7}. Multiple-Code (MC) and Multicode-Keying (MK) schemes are also proposed to support multirate transmission in OCDMA systems \cite{new_ref_10, new_ref_4}. Furthermore, some researchers have concentrated on designing new code families and efficient coding/decoding structures for OCDMA systems to reduce design complexity and increase system capabilities \cite{new_ref_3, new_ref_5, new_ref_6, new_ref_8, new_ref_11}. Performance analysis of different OCDMA-based systems is another attractive research topic \cite{new_ref_2, new_ref_9, new_ref_10, new_ref_12,ref_3, new_ref_13}. 

As examples of OCDMA-based systems, we can refer to the demonstrations of OCDMA Passive Optical Networks (OCDMA-PONs) for next generation high speed access networks \cite{new_ref_3, new_ref_14, ref_4, new_ref_15}. OCDMA-PON as a candidate for Next-Generation (NG)-PON can provide full asynchronous communication, data confidentiality, and symmetric bandwidth for downlink and uplink \cite{new_ref_3}. Noting the exponential growth of highly-dynamic traffic patterns due to emergence of social networking, real-time gaming and high definition audio-video streaming in modern networks, on-demand resource allocation is essential for cost reduction and revenue generation. Consequently, on-demand and agile resource provisioning in PONs is rapidly becoming a significant technical and economic priority. Software Defined Networking (SDN) as an emerging technology is a promising candidate for PON control plane to provide the favorable online resource assignment. Through the separation of control and data plane in network elements, SDN provides dynamic and fine-grained traffic control that enhances total controllability, manageability, efficiency and resource utilization of PONs  \cite{new_ref_1, new_ref_7}. Reducing the complexity of resource management algorithms is the main challenge that should be tackled to let an OCDMA-PON provide on-demand and adaptive resource allocation \cite{new_ref_1, new_ref_7, new_ref_16}. There is a strong desire to increase the bandwidth while decreasing the power consumption of NG-PON systems. Putting network element into sleep mode is a cost-effective and attractive method developed to decrease the power consumption in NG-PON. Performance of different green architectures for NG-PONs has been compared in \cite{new2_ref_8}. Many researchers have also proposed various green bandwidth allocation frameworks and algorithms for WDM- or TDM-PONs \cite{new2_ref_7, new2_ref_8, new2_ref_9} but green resource allocation for OCDMA-PON needs more research. 
 
Initially, OCDMA system was introduced by 1D Strict/Generalized Optical Orthogonal Codes (SOOC/GOOCs) as the user code strings.  2D wavelength-time optical coding schemes have also been studied for OCDMA system which improves overall network throughput and code cardinality \cite{ref_9,new_ref_5, new_ref_8, new2_ref_1}. Multi-class OCDMA systems have been proposed to provide different QoS levels that distinguish between users in terms of performance metrics such as transmission rate and BER. There are many techniques that can be used to design multi-class OCDMA systems \cite{ALC,ML, Div, MS}. System level concepts such as Multi-Length OCDMA (ML-OCDMA), Variable-Weight OCDMA (VW-OCDMA) and Multi-Length Variable-Weight OCDMA (MLVW-OCDMA) have been introduced for this purpose \cite{MS}. Differentiating in code parameters plays the key role in QoS provisioning of a MLVW-OCDMA system. Multi-level signaling technique has also been proposed in which multiple classes of QoS can be created by differentiating in transmitted power levels \cite{ML}. Diversity in OCDMA system can also be used to provide various levels of QoS differentiated by BER performance without changing the code parameters \cite{Div}. 

Simple and efficient dynamic resource allocation and system redesign is a key requirement for adaptive on-demand control of OCDMA systems. Since BER equations of OCDMA systems are not simple to calculate and are not easy to reverse, the redesign and dynamic resource assignment problems are consequently not easy to solve using few simple design equations and algorithms. In this paper, an approximated simple expression for the BER of a multi-class OCDMA system is derived and its accuracy is verified by comparing its results against the previously published BER equations. The proposed approximation is valid for OCDMA systems with 1D/2D optical codes. It also supports multi-class OCDMA systems having QoS classes differentiated by the mentioned system level techniques. The proposed BER equation provides insight into proper system level analysis, design and their corresponding sensitivity analysis. Considering code design, code cardinality and system performance constraints, we formulate two design problems that aim at maximizing transmission rate and minimizing optical power consumptions. We investigate heuristic and brute-force search approaches for the optimal solutions of the design problems and use them to provide design guidelines and procedures for a typical OCDMA system. The procedures are simple, easy to apply and can easily be used to adaptively derive the required system parameters during on-demand resource provisioning for dynamic traffic patterns. We also use the BER equation and design procedures to propose two simple code allocation algorithms, named "rate-efficient" and "power-efficient" code allocation algorithms, for an adaptive OCDMA-PON system that dynamically reallocate user codes according to the activity of users and dynamic traffic fluctuations. The rate-efficient code allocation algorithm exploits the unused resources of inactive users to increase the transmission rate of active ones while the power-efficient algorithm provides a green code allocation mechanism in which code weights are dynamically reduced to decrease optical power consumption. Simulation results show that the code allocation algorithms can use the randomness of the user activities to increase transmission rate or decrease power consumption and the amount of improvement is a function of system BER.

The rest of the paper is organized as follows. The system model and the methods of QoS provisioning are described in Section \ref{sec_II}. The approximated BER expression is derived and validated in Section \ref{sec_III}. In Sections \ref{sec_IV} and \ref{sec_V}, we show the applications of the proposed BER expression in developing systematic design and adaptive code allocation procedures. Simulation results are included in Section \ref{sec_VI}. Finally, we conclude the paper in Section \ref{sec_VII}.  
\section{System Model}\label{sec_II}
The system model considered in this paper is the common On-Off Keying (OOK) OCDMA with chip rate $R_c$ and $N$ active users \cite{new_ref_17,ref_9}. Each user has a transmitter that consists of an OOK modulator and a 1D/2D optical delay line encoder. A unique 2D optical code $(M L, W,\lambda)$ is assigned to each user where $M$, $L$, $W$ and $\lambda$ are number of wavelengths, code length, code weight and code cross-correlation, respectively. To guarantee the unique code assignment, the number of users $N$ should be less than code cardinality $\Phi(M,L, W,\lambda)$ which is in turn upper-limited by Johnson bound \cite{new2_ref_1, ref_15}:
\begin{equation}\label{jnb}
\Phi(M,L, W,\lambda) \leqslant \frac{M(ML-1)\cdots(ML-\lambda)}{W(W-1)\cdots(W-\lambda)}
\end{equation}
Transmitters send the assigned optical codes of their intended users for each $"1"$  data bit and no thing is sent during the bit interval of each $"0"$ data bit. Users are transmitting and receiving their chip sequences using optical fiber cables that are connected together using a passive star coupler. Transmitted signals are positively added in the star coupler and then a mixed signal is delivered to receivers. Receiver structure is based on the well-known optical AND logic gate receiver \cite{ref_9}. We assume that the main source of system performance degradation is MAI from other users and other system impairments such as background noise, shot noise and thermal noise are negligible. In fact, a practical OCDMA system can be considered in the MAI-limited regime if optical power is sufficiently higher than noise floor \cite{noise1}. The effects of noise in OCDMA systems are studied in many papers such as \cite{noise2} and can be combined with our MAI-limited analysis. 
  
Since all users of a conventional OCDMA system are given the same code parameters, power level and diversity order, they experience equal transmission rate and BER or in other words, the same QoS. In order to provide differentiated classes of QoS, ML-OCDMA has been introduced in which code words have the same code weight but different code length. VW-OCDMA is another multi-class OCDMA system in which code words have the same code length but different code weight. Multiple classes of QoS can also be provided using simultaneous differentiation in code weight and code length in MLVW-OCDMA system. In a MLVW-OCDMA system, the length and the weight of code words are chosen based on the desired QoS level such that high-weight and short-length code words are assigned to users having lower BER and higher transmission rate \cite{MS}. Multi-level signaling has been proposed to improve the performance of OCDMA systems and to provide multiple classes of QoS. In a multi-level OCDMA system, users are divided into different groups each transmitting at different power levels. In a multi-level OCDMA system, high-power classes have better BER performance than low-power ones\cite{ML}. Diversity in OCDMA deals with transmission of multiple copies of a bit through spatially independent optical paths and can improve BER performance without affecting code parameters. Diversity scheme proposed in \cite{Div} can be used to differentiate users in terms of their BER. We use the notation $\text{OCDMA}(K, M, \bold{N}, \bold{L}, \bold{W}, \bold{\Gamma}, \bold{C}, \bold{B})$ to characterize a general $K$-class $M$-wavelength 1D/2D OCDMA system where vectors $\bold{N}[N_i]_{1\times K}$, $\bold{L}[L_i]_{1\times K}$, $\bold{W}[W_i]_{1\times K}$ and $\bold{B}[b_i]_{1\times K}$ are number of users, code length, code weight and diversity order vectors, respectively. Matrices $\bold{\Gamma}[\lambda_{ij}]_{K\times K}$ and $\bold{C}[c_{ij}]_{K\times K}$ are code cross-correlation and power ratio matrices. Clearly, element $i$ of each vector corresponds to $i$-th class and $\lambda_{ij}$ denotes for cross-correlation between classes $i$ and $j$ code words. Furthermore, $c_{ij}$ shows the power ratio of class $i$ to class $j$ defined as: 
\begin{equation} 
c_{ij} = 
\begin{cases} 
\lfloor \frac{P_i}{P_j} \rfloor &\mbox{if } P_i \geqslant P_j \\ 
1 &\mbox{if } \hspace{2mm} P_i < P_j 
\end{cases} 
\end{equation} 
where $P_i$ and $P_j$ are the power levels of classes $i$ and $j$, respectively. Clearly, the transmission rate of $k$-th class in a multi-class OCDMA system is given by $R_k =\frac{R_c}{L_k}$ where $R_c$ is the mentioned common chip rate of the system.

Without loss of generality, we apply the concept of analysis and design of an OCDMA system for the case of an adaptive OCDMA-PON system as shown in Fig. 1. In this system, ONUs are generating their upstream data using an 1D/2D tunable transmitter in which the code parameters are programmable. As in other PON systems, all upstream data from ONUs are mixed together in the passive combiner and the combined signal reaches OLT. In OLT, a splitter generates $N$ version of the received signal which are fed to a bank of $N$ tunable 1D/2D receivers that decode the received upstream signal \cite{new2_ref_3,new2_ref_4}. We only consider MAI in our analysis and assume that ONU transmit powers are adjusted such that the received optical power from all ONUs are equal \cite{new2_ref_5}.
\begin{figure}
    \centering
    \begin{subfigure}[b]{0.5\textwidth}
        \includegraphics[width=\textwidth]{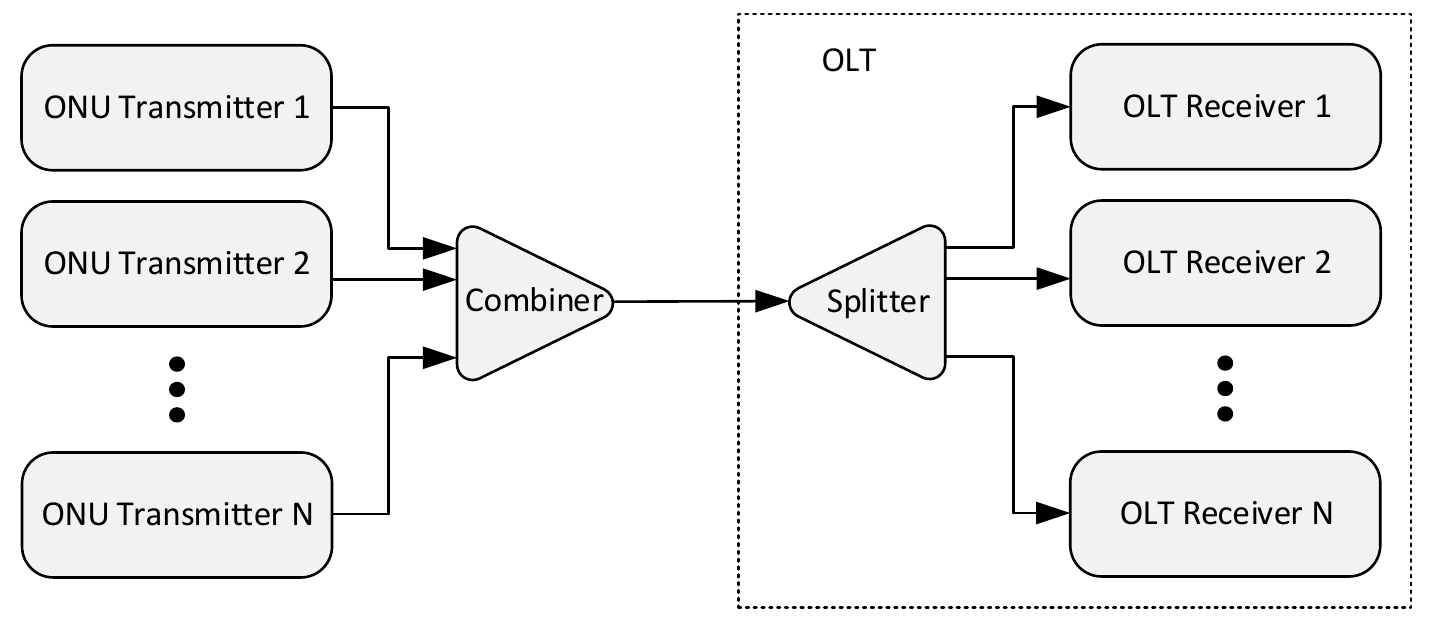}
        \caption{System block diagram}
        \label{pon_block}
    \end{subfigure}
    ~ 
    \begin{subfigure}[b]{0.5\textwidth}
        \includegraphics[width=\textwidth]{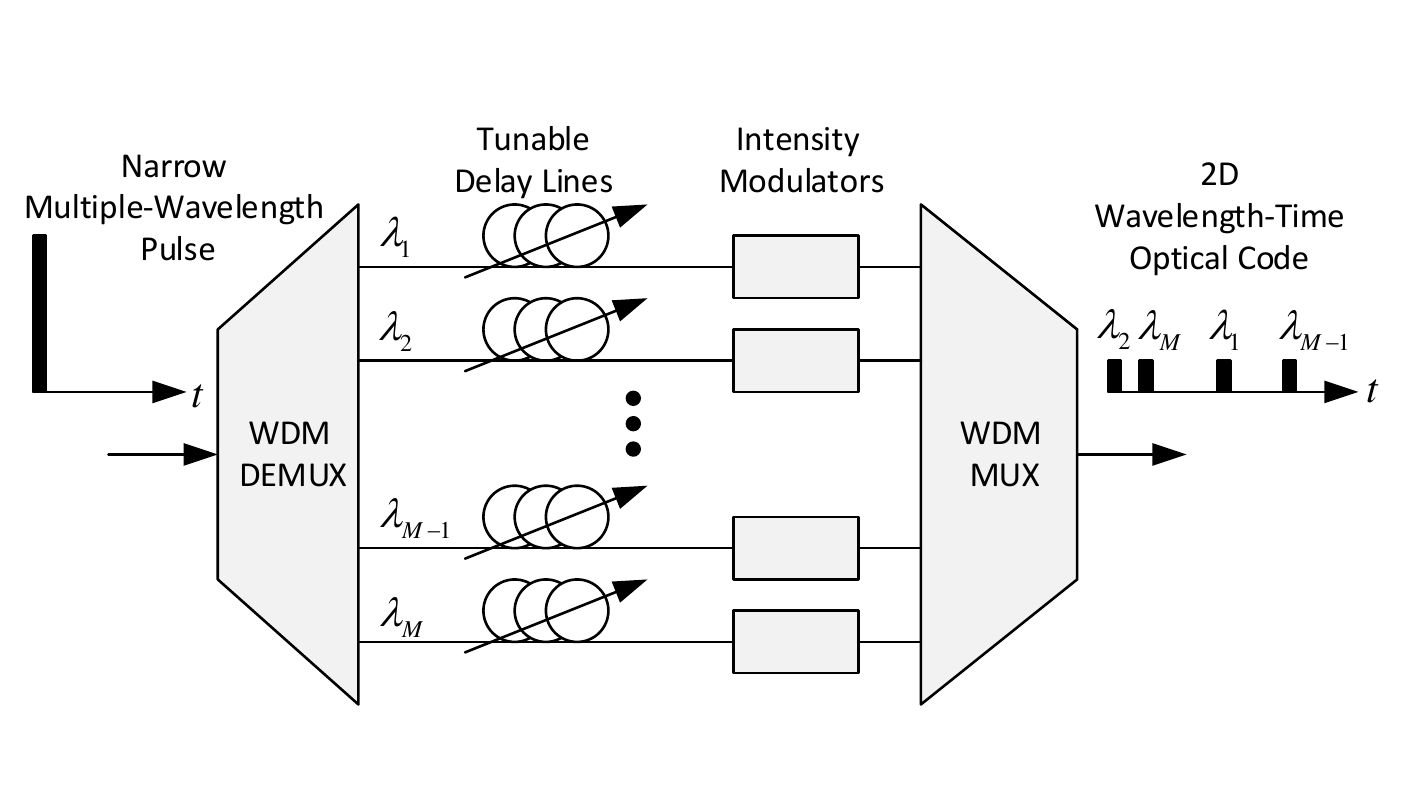}
        \caption{ONU Transmitter}
        \label{pon_tx}
    \end{subfigure}
    ~ 
    \begin{subfigure}[b]{0.5\textwidth}
        \includegraphics[width=\textwidth]{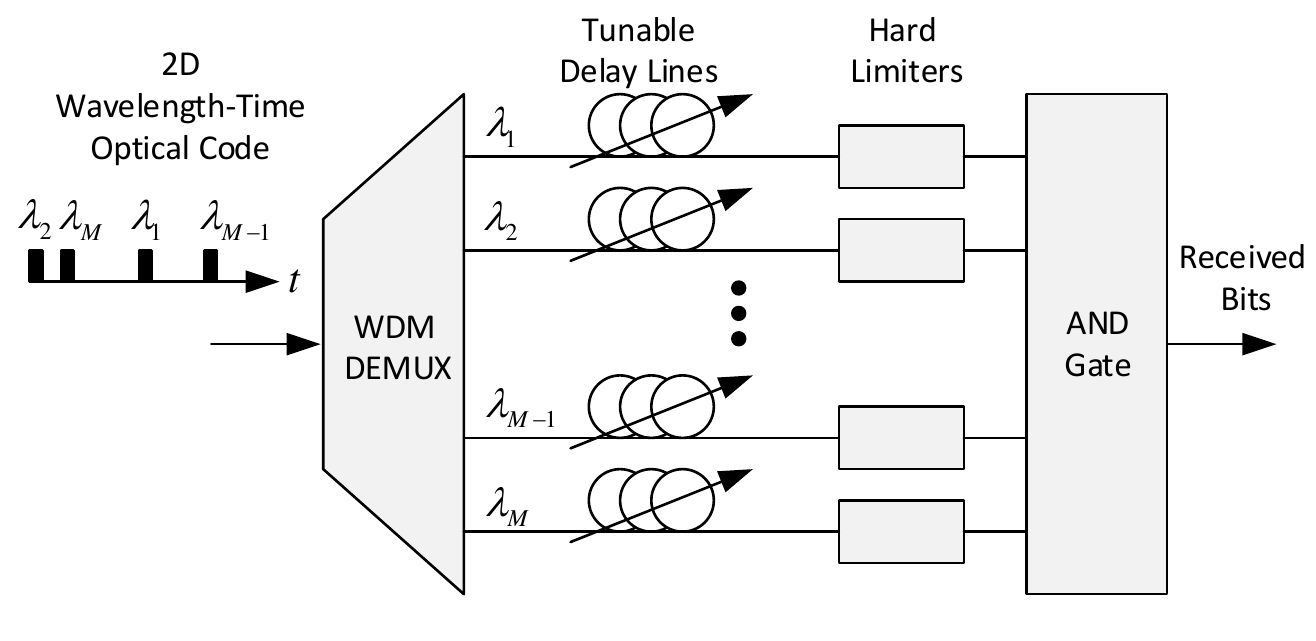}
        \caption{OLT Receiver}
        \label{pon_rx}
    \end{subfigure}
    \caption{\label{pon}Adaptive OCDMA-PON system block diagram in upstream which includes ONU transmitters, OLT receivers, optical combiner and optical splitter. Block diagram of the ONU transmitter and OLT receiver are also shown in which tunable optical delay lines can be adjusted to fit a desired optical code.}
\end{figure}
\section{Approximated BER Expression}\label{sec_III}
\begin{figure}
\center{\includegraphics[scale=0.47]{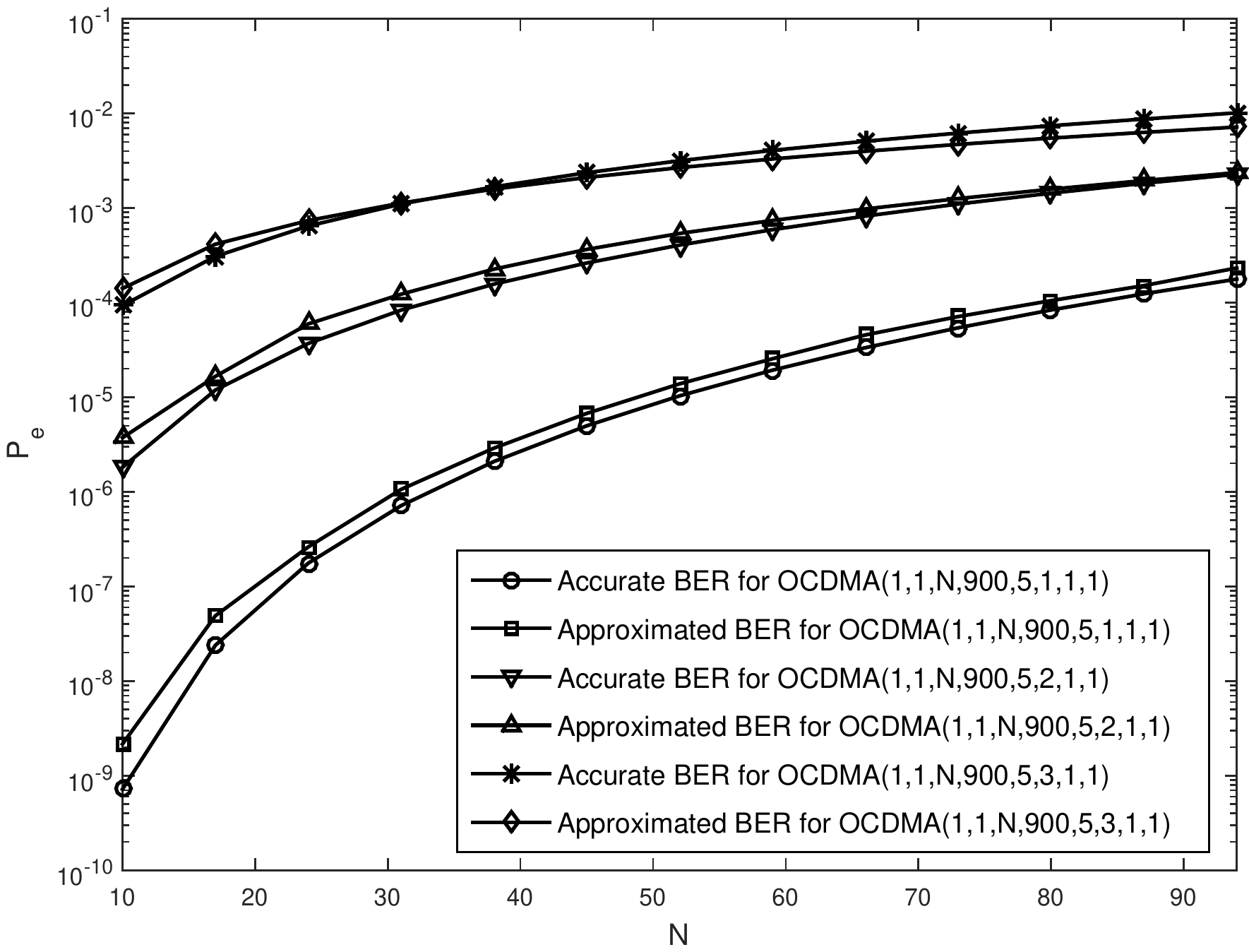}}
\center{\caption{\label{sim} Comparison of accurate BER and its approximation in terms of the number of users for three single-class 1D OCDMA systems distinguished by different values of cross-correlation $\lambda = 1, 2, 3$.  In this comparison, we assume that $P_{m}=0, m = 1, \cdots, \lambda-1$ to compare the worst case BER performance.}}
\end{figure}

The BER of $k$-th class in an OCDMA system characterized by $\text{OCDMA}(K, 1, \bold{N}, \bold{L}, \bold{W}, \bold{\Gamma}, \bold{1}, \bold{1})$ is given by\cite{MS}:
\begin{align}\label{eq_4}
& P_{e(k)} =\frac{1}{2} \sum\limits_{i=0}^{W_k} \bigg\{ (-1)^i {W_k \choose i}\\
\nonumber  & \prod\limits_{q=1}^{K}\Big[1+\sum\limits_{j=1}^{\lambda_{kq}}\sum\limits_{m=j}^{\lambda_{kq}}(-1)^j {i \choose j}{W_k - m\choose m-j}P_{m}^{kq}\Big]^{N_q^{k}} \bigg\}
\end{align}
where $N^k_q$ is the number of interfering users with values $N^k_q = N_q, q \neq k$ and $N^k_k = N_k-1$. $P_{m}^{kq}$ is the probability that a class $q$ user makes $m$ interferences on a class $k$ user. Obviously, \eqref{eq_4} is a complicated function of code parameters and number of users. Furthermore, 2D, multi-level and diversity signaling severely increase the complexity of BER expression \cite{ML, Div}. To relax the inherent complexity of the BER expression, we have derived a simple form approximation that can provide accurate estimation of the BER value for a general multi-class OCDMA system:
\begin{align}\label{appber}
P_{e(k)} \approx \frac{1}{2}\bigg[\sum\limits_{i=1}^{K}\Big(\frac{N_iW_i}{2M\lambda_{ki} L_i}\Big)^{\frac{c_{ki}}{\lambda_{ki}}}\bigg]^{W_kb_k}
\end{align}
The method of deriving the approximated BER expression is included in the Appendix. For a single class OCDMA, the approximated BER expression is simplified to:
\begin{equation}\label{appber1c}
P_e \approx \frac{1}{2}(\frac{NW}{2ML\lambda})^{\frac{W}{\lambda}}
\end{equation}
As a special case, for a single class OCDMA system with SOOC, the approximation is reduced to $P_e \approx \frac{1}{2}(\frac{NW}{2ML})^W$ which is equal to the approximation proposed in \cite{Div}. The accuracy of the approximation can be verified by comparing its values against the values offered by previously published BER equations \cite{ML, Div,MS}. As example,  Fig. \ref{sim} compares the accurate BER and its approximation in terms of the number of users for three single-class 1D OCDMA systems distinguished by different values of cross-correlation $\lambda = 1, 2, 3$.
\section{Systematic Design Procedures}\label{sec_IV}
In this part, we formulate two design problems to find the optimum values of code parameters that result in maximum transmission rate or minimum optical power consumption. We refer to these formulations as "Rate-optimized" and "Power-optimized" design problems.
\subsection{Rate-Optimized Design Problem}
Assume an 1D/2D OCDMA-PON system having chip rate $R_c$, $N$ users, $M$ wavelengths using GOOC codes.. We want to determine the values of $L$, $W$ and $\lambda$ such that code design and cardinality constraints are met, BER is less than a predefined value $P_{e_{th}}$ and transmission rate is maximized (or equivalently code length $L$ is minimized):
\begin{align}\label{rateopt}
& \min \hspace{4 mm} R \\
\nonumber & s.t. 
\begin{cases} 
C1:\hspace{2mm}\lambda \leqslant W \leqslant ML\\ 
C2:\hspace{2mm}NW(W-1)\cdots(W-\lambda) \leqslant ML(ML-1)\cdots(ML-\lambda)\\ 
C3:\hspace{2mm}W \leqslant \sqrt{2M\lambda L}\\
C4:\hspace{2mm}P_e(1, M, N, L, W, \lambda,1, 1) \leqslant P_{e_{th}}\\
C5:\hspace{2mm}W,L,\lambda \in \mathbb{N}
\end{cases} 
\end{align}
$C1$ is an obvious constraint on optical code deign and $C2$ comes from cardinality constraint \eqref{jnb}. $C3$ is an extension of the proposed bound in \cite{ref_9} to 2D optical codes. Finally, $C4$ specifies BER performance requirement. Brute force search over 3D region limited to $W_{max}$, $L_{max}$ and $\lambda_{max}$ needs $W_{max}L_{max}\lambda_{max}$ evaluations of constraints which may be complex and unacceptable for online redesign or reoptimization of the system. We provide a heuristic design routine shown in Alg. \ref{flowchart} which finds the solution in a much simpler and faster method. For fixed value of $\lambda$ over continuous 2D region of $L$ and $W$, $C1$, $C2$ and $C3$ yield ascending boundary functions for $L$ in terms of $W$ while $C4$ provides a descending one (remember that for a fixed value of $P_{e_{th}}$, $L$ is a descending function of $W$). Therefore, the optimum point is certainly placed on the intersections of $C4$ boundary with boundaries $C1$, $C2$ and $C3$. This implies that for the discrete case, the optimum solution is around these intersections. As can be seen from \eqref{appber1c}, BER is exponentially increased for higher values of $\lambda$ and as code cardinality constraint $C2$ allows we should choose the smallest $\lambda$. In each iteration of Alg. \ref{flowchart}, a limited region around the mentioned intersections is searched for the smallest $L$ and the algorithm stops when the smallest $\lambda$ and $L$ are obtained. Note that we approximate $C2$ by $LM\geqslant \sqrt[\leftroot{-3}\uproot{3}\lambda]{NW(W-1)\cdots(W-\lambda)/M}$ to obtain an explicit boundary expression for $L$. In addition, one can use the approximated BER \eqref{appber1c} to provide an initial value to speed up local searches around the intersections. 
\begin{algorithm}[t!]
\caption{Heuristic procedure for designing an OCDMA-PON.}\label{flowchart}
\begin{algorithmic}[1]
\Input{Number of users $N$, Number of wavelength $M$, Error threshold $P_{e_{th}}$}
\Output{Code Length $L$,  Code weight $W$, Code cross-correlation $\lambda$}
\Hline{\hspace{-.6 cm}\hrulefill}
\State $\lambda \longleftarrow 1$;
\State search for the smallest $W$ such that $P_e(1, M, N, L, W, \lambda,1, 1) \leqslant P_{e_{th}}$ where $L = \frac{NW(W-1)+M}{M^2}^\dagger$;
\While{1}
\State $\lambda_t \longleftarrow  \lambda + 1$;
\State search for the smallest $W_{t1}$ such that 
\NoNumber{$P_e(1, M, N, L_{t1}, W_{t1}, \lambda,1, 1) \leqslant P_{e_{th}}$ where}
\NoNumber{$L_{t1} = \frac{1}{M}\sqrt[\leftroot{-3}\uproot{3}\lambda_t]{NW_{t1}\cdots(W_{t1}-\lambda_t)/M} \pm j, j = 1,\cdots, \lambda_t^\dagger$;}
\If{$\frac{W_{t1}^2}{2M\lambda_t} > L_{t1}$}
\State $L_{t1} \longleftarrow \infty$;
\EndIf
\State search for the smallest $W_{t2}$ such that 
\NoNumber{$P _e(1, M, N, L_{t2}, W_{t2}, \lambda,1, 1) \leqslant P_{e_{th}}$ where $L_{t2} = \frac{W_{t2}^2}{2M\lambda_t}^\dagger$;}
\If{$N > \frac{M(ML_{t2}-1)\cdots(ML_{t2}-\lambda_t)}{W_{t2}(W_{t2}-1)\cdots(W_{t2}-\lambda_t)}$}
\State $L_{t2} \longleftarrow \infty$;
\EndIf
\If{$\min\{L_{t1}, L_{t2}\} < L$}
\State $\lambda \longleftarrow \lambda_t$;
\If{$L_{t1} < L_{t2}$}
\State $W \longleftarrow W_{t1}$, $L \longleftarrow  L_{t1}$;
\Else
\State $W \longleftarrow  W_{t2}$ , $L \longleftarrow  L_{t2}$;
\EndIf
\Else
\State \textbf{break};
\EndIf
\EndWhile
\State store $L$, $W$, $\lambda$; 
\end{algorithmic}
\footnotesize{$^\dagger$ use approximation \eqref{appber1c} to choose an initial and speed up local search.}
\end{algorithm}
\subsection{Power-Optimized Design Problem}
Again consider the 1D/2D OCDMA-PON system with chip rate $R_c$, $N$ users and $M$ wavelengths with GOOCs and now assume that code length $L$ (equivalently transmission rate $R$) is given. Clearly for fixed values of $L$ and $M$, average optical power is proportional to code weight $W$. We want to find a pair of $W$ and $\lambda$ that minimizes optical power (or equivalently minimizes code weight) and holds BER, code design and code cardinality constraints. The problem is formulated as below:
\begin{align}\label{powopt}
& \min \hspace{4 mm} W\\
\nonumber & s.t. 
\begin{cases} 
C1:\hspace{2mm}\lambda \leqslant W \leqslant ML\\ 
C2:\hspace{2mm}NW(W-1)\cdots(W-\lambda) \leqslant ML(ML-1)\cdots(ML-\lambda)\\ 
C3:\hspace{2mm}W \leqslant \sqrt{2M\lambda L}\\
C4:\hspace{2mm}P_e(1, M, N, L, W, \lambda,1, 1) \leqslant P_{e_{th}}\\
C5:\hspace{2mm}W,\lambda \in \mathbb{N}
\end{cases} 
\end{align}
A brute force search over 2D region limited to $W_{max}$ and $\lambda_{max}$ can yield the optimal solution of \eqref{powopt} by $W_{max}\lambda_{max}$ evaluation of constraints. Since the number of search points is small, it is easy to find the optimum point using a simple search process.
\section{Adaptive Code Allocation Algorithms}\label{sec_V}
Here, we design an OCDMA-PON using the introduced design procedures and then propose two simple and adaptive code allocation algorithms that dynamically reallocate user codes to exploit the unused resources of inactive users. The first algorithm shares the unused resources of the inactive users among active ones to increase the transmission rate while the other algorithm adaptively reduces the code weight to decrease the amount of optical power consumption when the number of inactive users falls in predefined integer intervals. 
\subsection{Rate-Efficient Adaptive Code Allocation Algorithm}
\begin{figure}
\center{\includegraphics[scale=0.6]{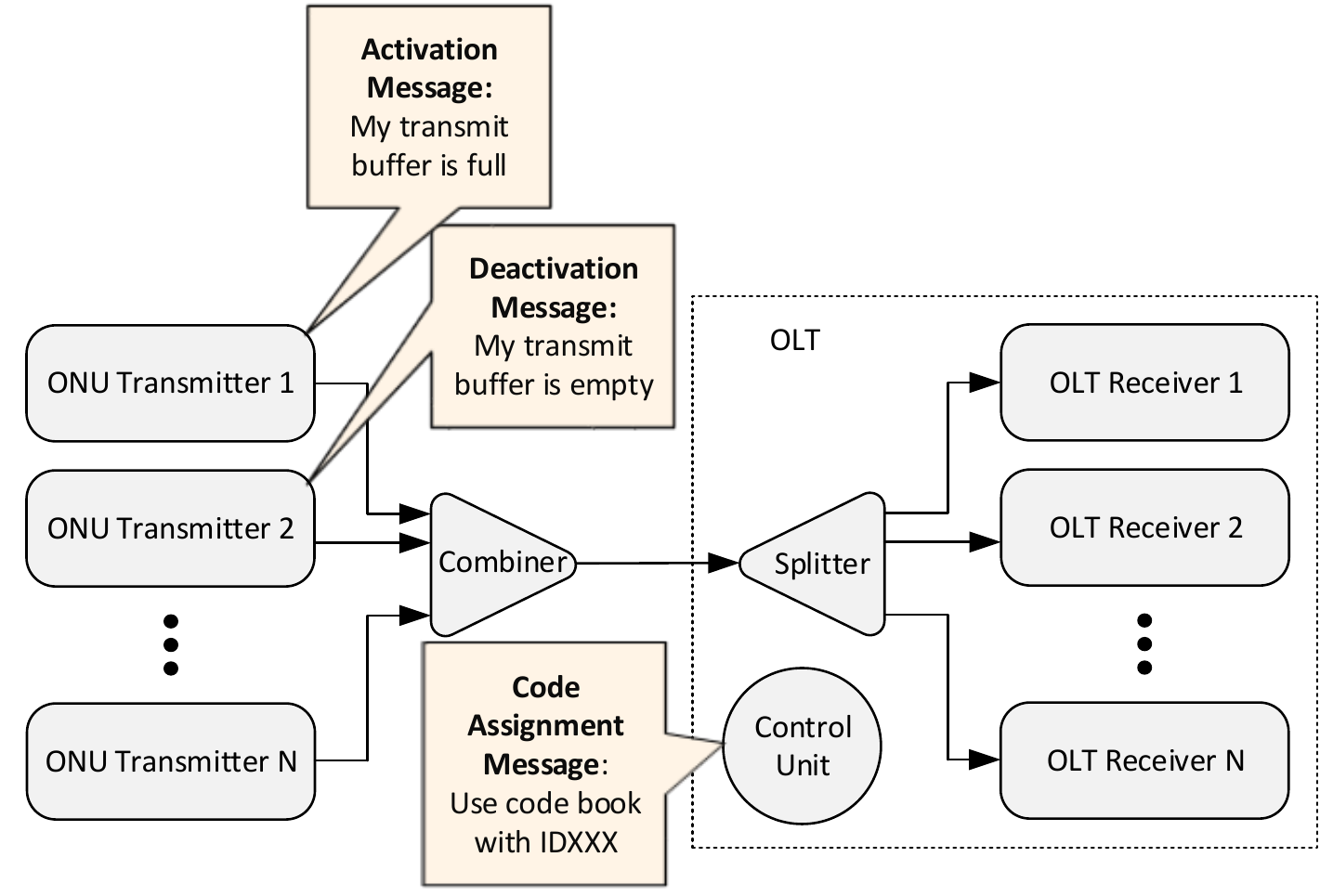}}
\center{\caption{\label{sdn_fig} Control unit and messages in the proposed adaptive OCDMA-PON system running algorithm Alg. \ref{flowchart2} or Alg. \ref{flowchart3}. }}
\end{figure}

Consider a single class 1D/2D OCDMA-PON system that works with $M$ wavelengths and GOOCs, has at most $N$ active users and its BER is upper-limited by $P_{e_{th}}$. Assume that the system has a bank of optical codes that includes different code books corresponding to various code parameters $L$, $W$ and $\lambda$. Each code book has a unique ID number. There are many methods that can be used to construct the code bank \cite{new_ref_20, ref_15}. Based on \eqref{appber1c}, BER approximately remains unchanged while the ratio $\frac{N}{L}$ is fixed which implies that code length and consequently transmission rate can adaptively be improved according to the number of active users. In our rate-efficient OCDMA-PON, the control unit which resides in OLT, dynamically monitors the system nodes, receives control messages, runs the proposed routine in Alg. \ref{flowchart2} and allocates available resources among the active users as illustrated in Fig. \ref{sdn_fig}. The algorithm has an offline stage in which the optimum values of code parameters for maximizing transmission rate in an OCDMA-PON with $i, i = 1, \cdots, N$ active users are computed using formulation \eqref{rateopt} and heuristic Alg. \ref{flowchart}. In online stage, number of active users is updated by means of received activation and deactivation messages and at predefined time interrupts separated by code reallocation time $T$, a new code book corresponding to the number of active users is selected and announced. Next, optical decoders and encoders are tuned according to the announced code assignment messages. 
\begin{algorithm}[t!]
\caption{Rate-efficient adaptive code allocation algorithm for OCDMA-PON system.}\label{flowchart2}
\begin{algorithmic}[1]
\Input{User activity information, Number of wavelength $M$, Maximum number of active users $N$, error threshold $P_{e_{th}}$, code reallocation interval $T$}
\Output{Code assignment messages}
\Hline{\hspace{-.6 cm}\hrulefill}
\For{$i=1, \cdots, N$}
\State design an OCDMA system with $i$ users, $M$ 
\NoNumber{wavelengths and error threshold $P_{e_{th}}$ using}
\NoNumber{formulation \eqref{rateopt} and name obtained code length,}
\NoNumber{code weight and cross-correlation $L_i$, $W_i$ and $\lambda_i$;}
\EndFor
\State $n \longleftarrow N$;
\State choose code book $(L_n,W_n, \lambda_n)$;
\State send code assignment messages;
\State $T_{cr} \longleftarrow \text{time()} + T$;
\While{$1$}
\If{a user activation message is received}
\State $n \longleftarrow n + 1$;
\ElsIf{a user deactivation message is received}
\State $n \longleftarrow n - 1$;
\EndIf
\If{$\text{time()} \geqslant T_{cr}$}
\State choose code book $(L_n,W_n, \lambda_n)$;
\State send code assignment messages;
\State $T_{cr} \longleftarrow \text{time()} + T$;
\EndIf
\EndWhile
\end{algorithmic}
\end{algorithm}
\subsection{Power-Efficient Adaptive Code Allocation Algorithm}
Again, assume that we have a single class 1D/2D OCDMA-PON with $M$ wavelengths, at most $N$ users and BER threshold $P_{e_{th}}$. If an OCDMA transmitter uses a code with smaller code weight $W$, it consumes less energy.  As can be seen from \eqref{appber1c}, if the number of active users is sufficiently low, transmitted optical power can be reduced by reallocating a new code book with lower code weight such that BER remains less than the target value. In offline phase of Alg. \ref{flowchart3}, an OCDMA system with given parameters $N$, $M$ and $P_{e_{th}}$ are design using heuristic Alg. \ref{flowchart}. Then, formulation \eqref{powopt} is used to find optimum values of $W$ and $\lambda$ for a fixed value of $L$ and various values of number of users $i, i = 1, \cdots, N$. Similarly in the online stage, number of active users is determined using activation and deactivation messages and at certain time interrupts, a new code book that corresponds to the number of active users is selected and announced. Fig. \ref{sdn_fig} illustrates the adaptive OCDMA-PON, its control unit and control messages.
\begin{algorithm}[t!]
\caption{Power-efficient adaptive code allocation algorithm for OCDMA-PON system.}\label{flowchart3}
\begin{algorithmic}[1]
\Input{User activity information, Number of wavelength $M$, Maximum number of active users $N$, error threshold $P_{e_{th}}$, code reallocation interval $T$}
\Output{Code assignment messages}
\Hline{\hspace{-.6 cm}\hrulefill}
\State design an OCDMA system with $N$ users, $M$ wavelengths and error threshold $P_{e_{th}}$ using formulation \eqref{rateopt} and name obtained code length, code weight and cross-correlation $L$, $W$ and $\lambda$;
\For{$i=1, \cdots, N$}
\State design an OCDMA system with $i$ users, $M$ 
\NoNumber{wavelengths, code length $L$ and error threshold $P_{e_{th}}$}
\NoNumber{using formulation \eqref{powopt} and name obtained code}
\NoNumber{weight and cross-correlation $W_i$ and $\lambda_i$;}
\EndFor
\State $n \longleftarrow N$;
\State choose code book $(L,W_n, \lambda_n)$;
\State send code assignment messages;
\State $T_{cr} \longleftarrow \text{time()} + T$;
\While{$1$}
\If{a user activation message is received}
\State $n \longleftarrow n + 1$;
\ElsIf{a user deactivation message is received}
\State $n \longleftarrow n - 1$;
\EndIf
\If{$\text{time()} \geqslant T_{cr}$}
\State choose code book $(L,W_n, \lambda_n)$;
\State send code assignment messages;
\State $T_{cr} \longleftarrow \text{time()} + T$;
\EndIf
\EndWhile
\end{algorithmic}
\end{algorithm}
\section{Simulation Results}\label{sec_VI}
\begin{figure}
\center{\includegraphics[scale=0.47]{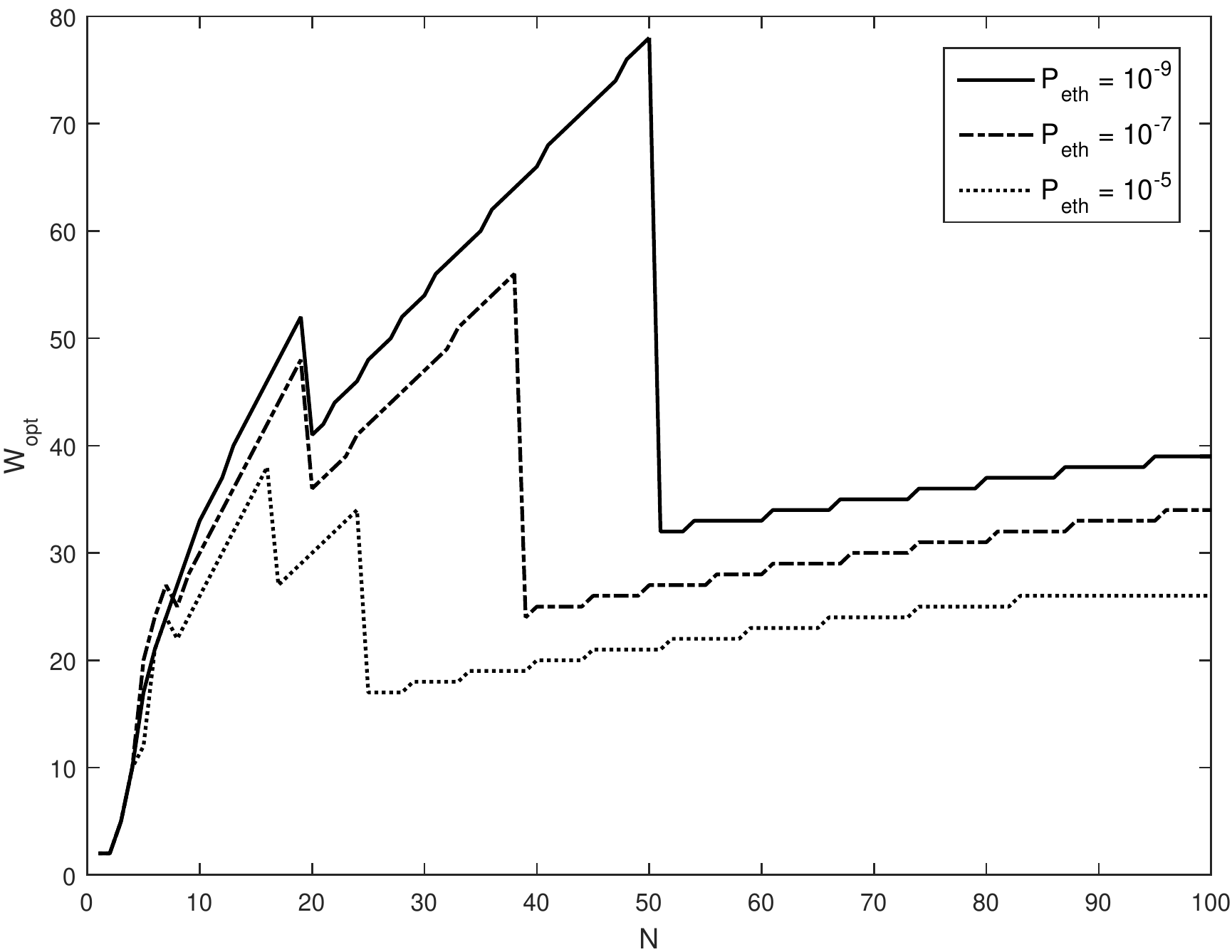}}
\center{\caption{\label{rateW} Optimum value of $W$ in a rate-optimized 1D OCDMA system design for various $P_{e_{th}}=10^{-5}, 10^{-7}, 10^{-9}$ versus number of users $N$. Alg. \ref{flowchart}  and brute force approaches yield the same results.}}
\end{figure}

\begin{figure}
\center{\includegraphics[scale=0.47]{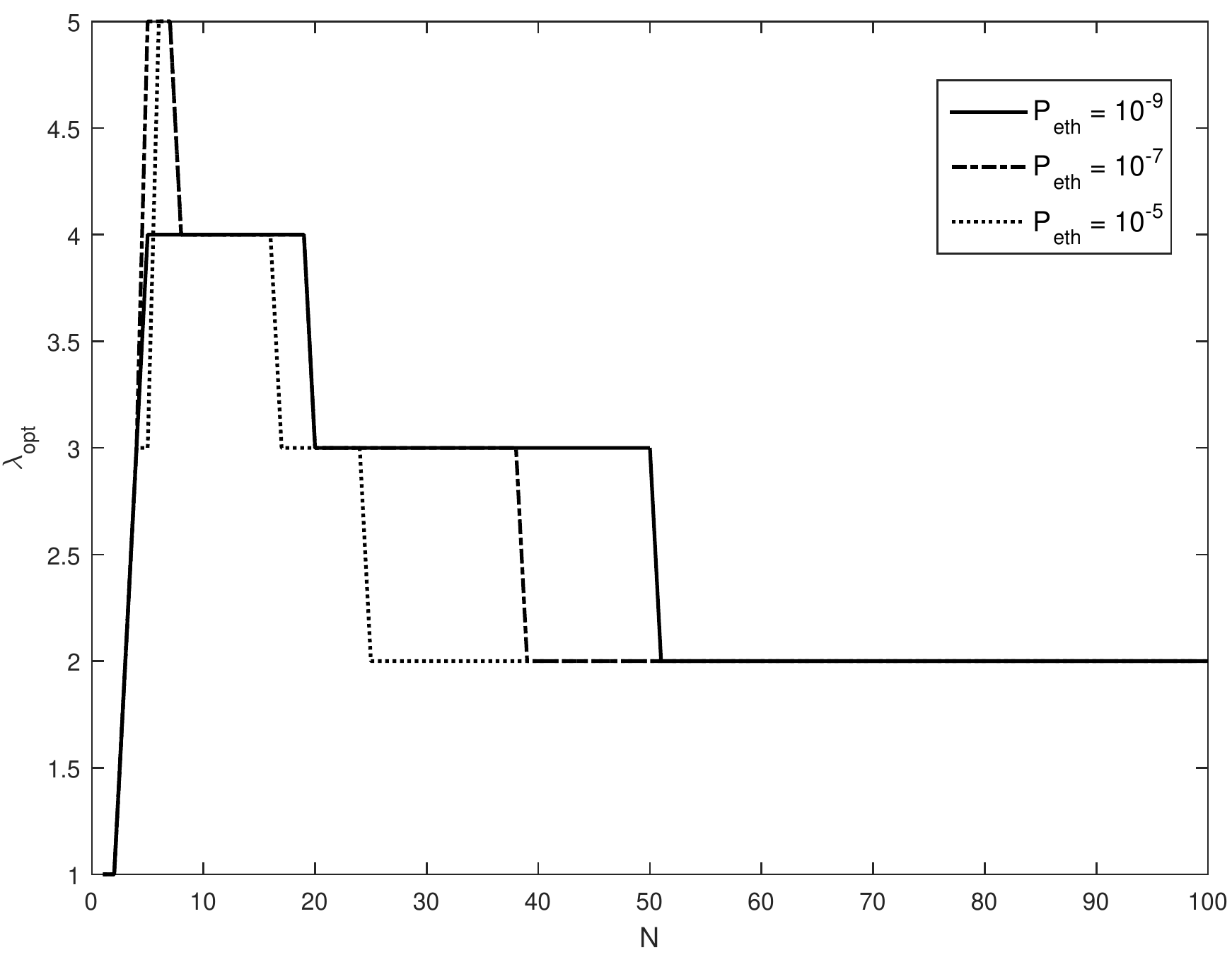}}
\center{\caption{\label{rateLam} Optimum value of $\lambda$ in a rate-optimized 1D OCDMA system design for various $P_{e_{th}}=10^{-5}, 10^{-7}, 10^{-9}$ versus number of users $N$. Alg. \ref{flowchart} and brute force approaches yield the same results.}}
\end{figure}

\begin{figure}
\center{\includegraphics[scale=0.47]{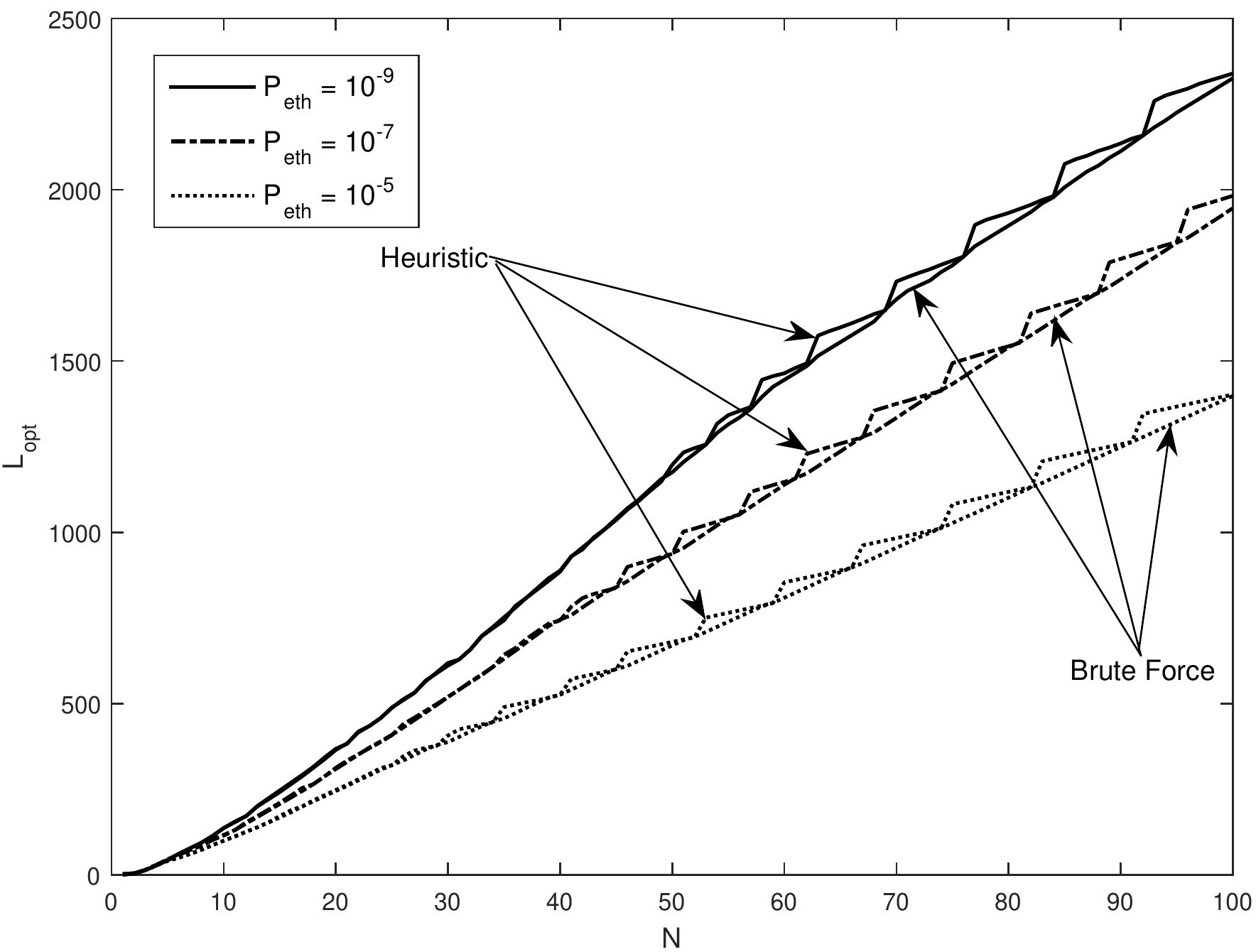}}
\center{\caption{\label{rateL} Optimum value of $L$ in a brute force rate-optimized 1D OCDMA system design for various $P_{e_{th}}=10^{-5}, 10^{-7}, 10^{-9}$ versus number of users $N$. For large $N$, estimated $L$ by Alg. \ref{flowchart} is slightly greater than the optimum one.}}
\end{figure}

\begin{figure}
\center{\includegraphics[scale=0.47]{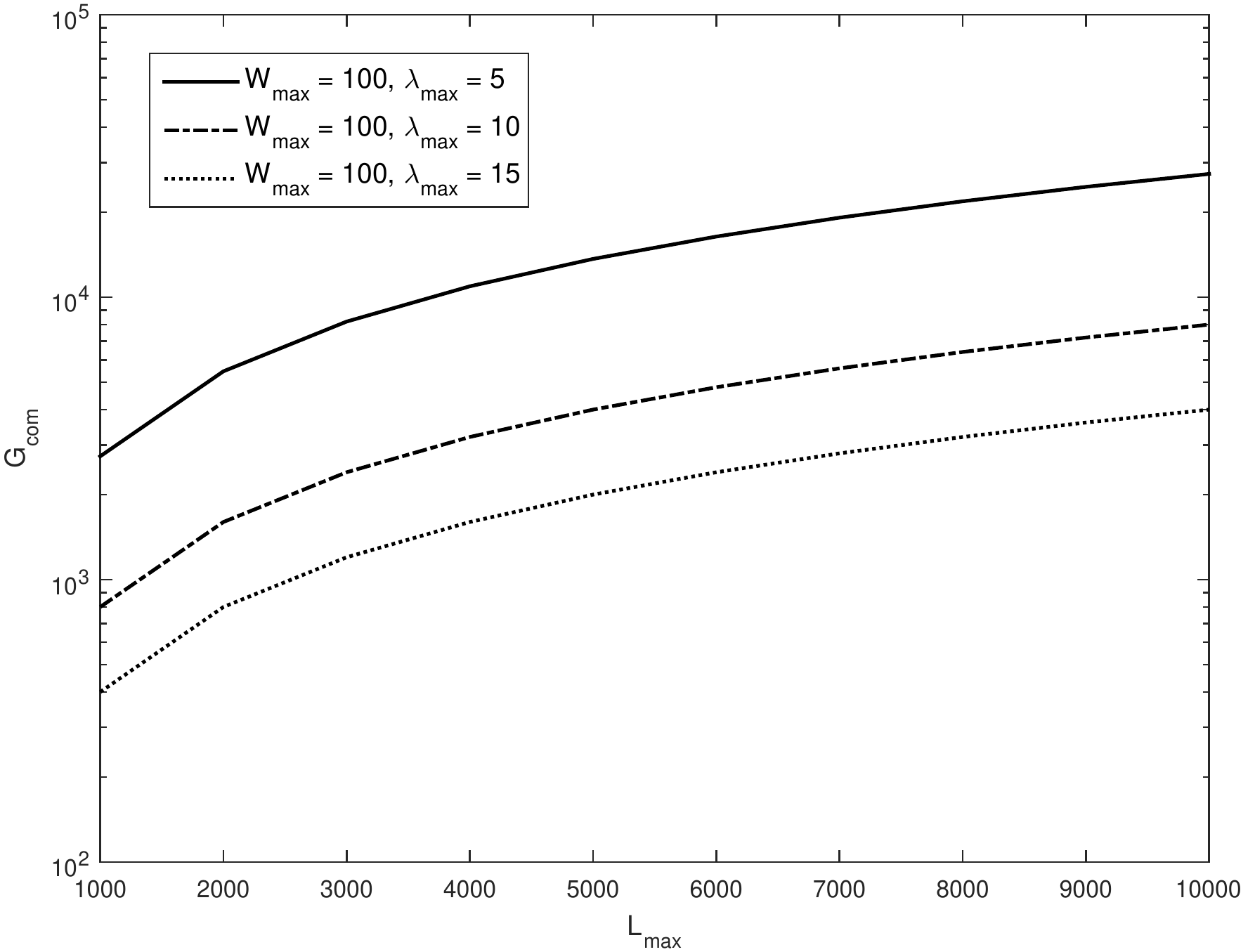}}
\center{\caption{\label{complexity} Complexity gain $G_{com}$ which is defined as the relative number of required MAC operations in rate-optimized brute force search over 3D region limited to $W_{max}$, $L_{max}$ and $\lambda_{max}$ to the required ones in Alg. \ref{flowchart}.}}
\end{figure}

Without loss of generality and for the sake of simplicity, we restrict our simulations to a 1D OCDMA system. To evaluate the performance of Alg. \ref{flowchart}, its results are compared with brute force search. Fig. \ref{rateW}, Fig. \ref{rateLam} and Fig. \ref{rateL} show the optimum values of $W$, $\lambda$ and $L$ in a 1D OCDMA system design for various $P_{e_{th}}=10^{-5}, 10^{-7}, 10^{-9}$ versus number of users $N$. As shown in the figures, there is no difference between the results of Alg. \ref{flowchart} and brute force search approaches for $W$ and $\lambda$. For large values of $N$, the offered $L$ by Alg. \ref{flowchart} is slightly greater than the optimum one obtained by the brute force search. Our simulation results show that $\lambda$ mainly takes $2$ or $3$ as its optimum values which is the same as the results published in \cite{ref_9}. However, for lower values of $N$, the optimum value of $\lambda$ may be greater than $3$. Fig. \ref{complexity} shows the amount of complexity reduction for Alg. \ref{flowchart} compared to the brute force search over the region limited to $W_{max}$, $\lambda_{max}$ and $L_{max}$. The vertical axis is complexity gain $G_{com}$ which is defined as the relative number of required MAC operations in the brute force search to the MAC operations of Alg. \ref{flowchart}. As shown in Fig. \ref{complexity}, Alg. \ref{flowchart} reduces the computational complexity, which is an important factor for on-demand and adaptive system design and resource management. For example when $L_{max} = 4000$, $W_{max} = 100$ and $\lambda_{max} = 5$, the complexity gain $G_{com}$ is more than $10^4$ which means that Alg. \ref{flowchart} is $10^4$ times faster than its brute force search counterpart.

\begin{figure}
\center{\includegraphics[scale=0.47]{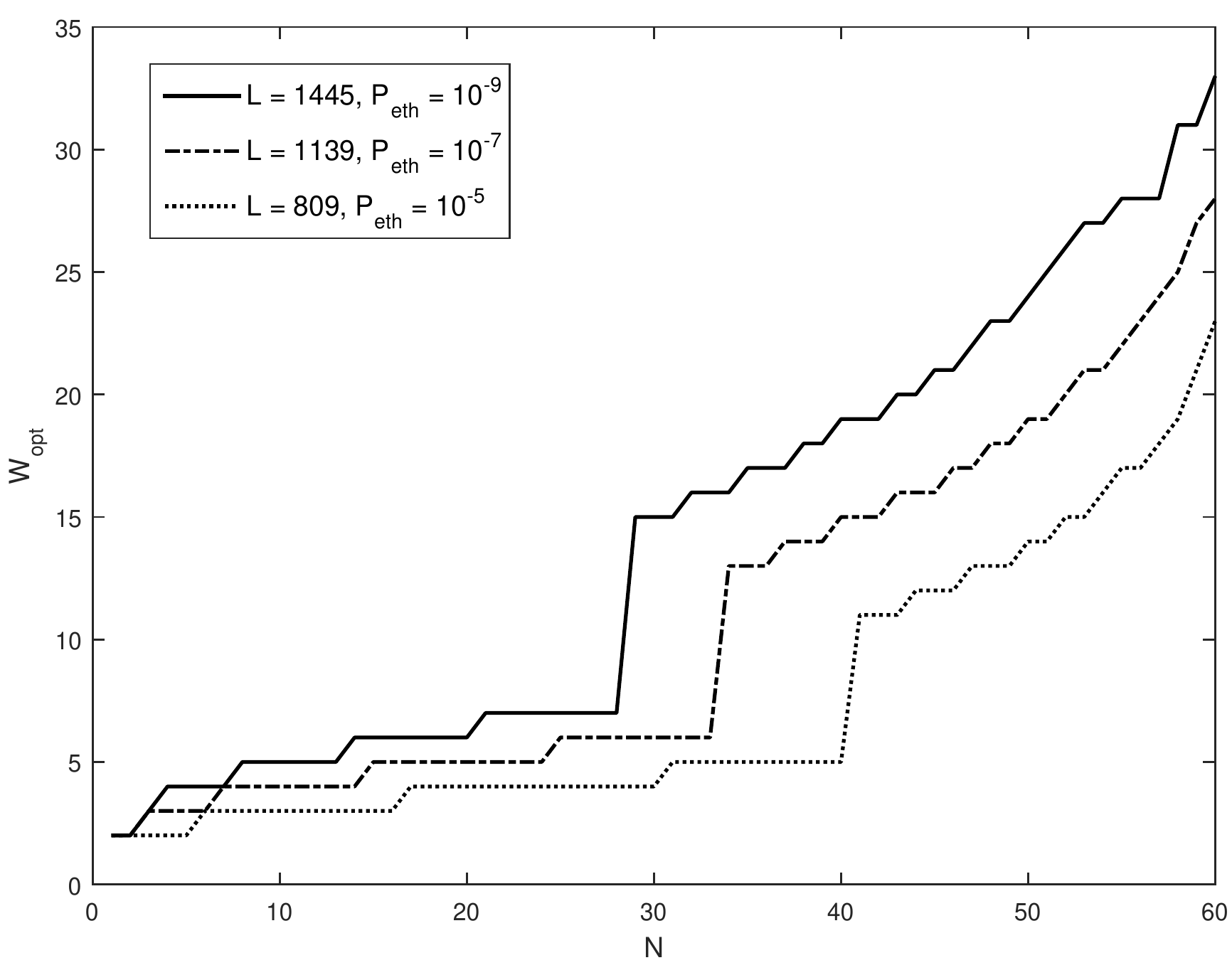}}
\center{\caption{\label{powW} Optimum value of $W$ in power-optimized 1D OCDMA system design versus number of users $N$ for fixed values of $L$ and various error thresholds $P_{e_{th}}$.}}
\end{figure}

\begin{figure}
\center{\includegraphics[scale=0.47]{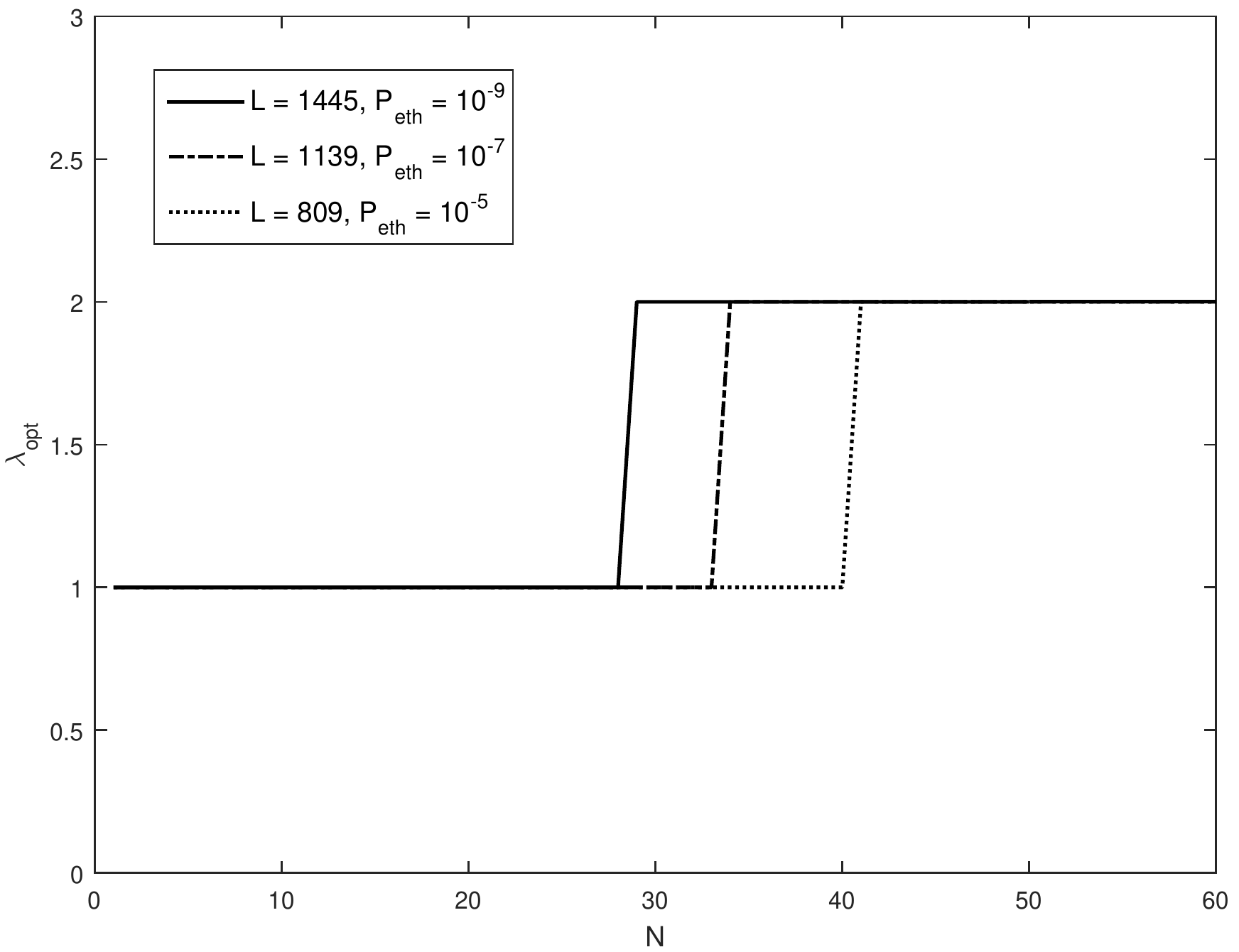}}
\center{\caption{\label{powLam} Optimum value of $\lambda$ in power-optimized 1D OCDMA system design versus number of users $N$ for fixed values of $L$ and various error thresholds $P_{e_{th}}$.}}
\end{figure}

The optimum values of $W$ and $\lambda$ versus number of users $N$ for power-optimized problem are plotted in Fig. \ref{powW} and Fig. \ref{powLam}, respectively. To obtain these figures, we first design three OCDMA systems with $N=60$ and $P_{e_{th}}=10^{-5}, 10^{-7}, 10^{-9}$ using the rate-optimized formulation \eqref{rateopt} to obtain code lengths which are $L=809, 1139, 1445$. Then, power-optimized brute force search for $N=1,\cdots,60$ and the obtained values of $L$ is executed. Note that the jumps for $W$ in Fig. \ref{rateW} and Fig. \ref{powW} correspond to points in which $\lambda$ changes. Based on \eqref{appber1c}, the exponent $\frac{W}{\lambda}$ has a significant effect on BER values and for any change in $\lambda$, $W$ can change such that $\frac{W}{\lambda}$ and consequently BER approximation remain constant.

\begin{figure}
\center{\includegraphics[scale=0.47]{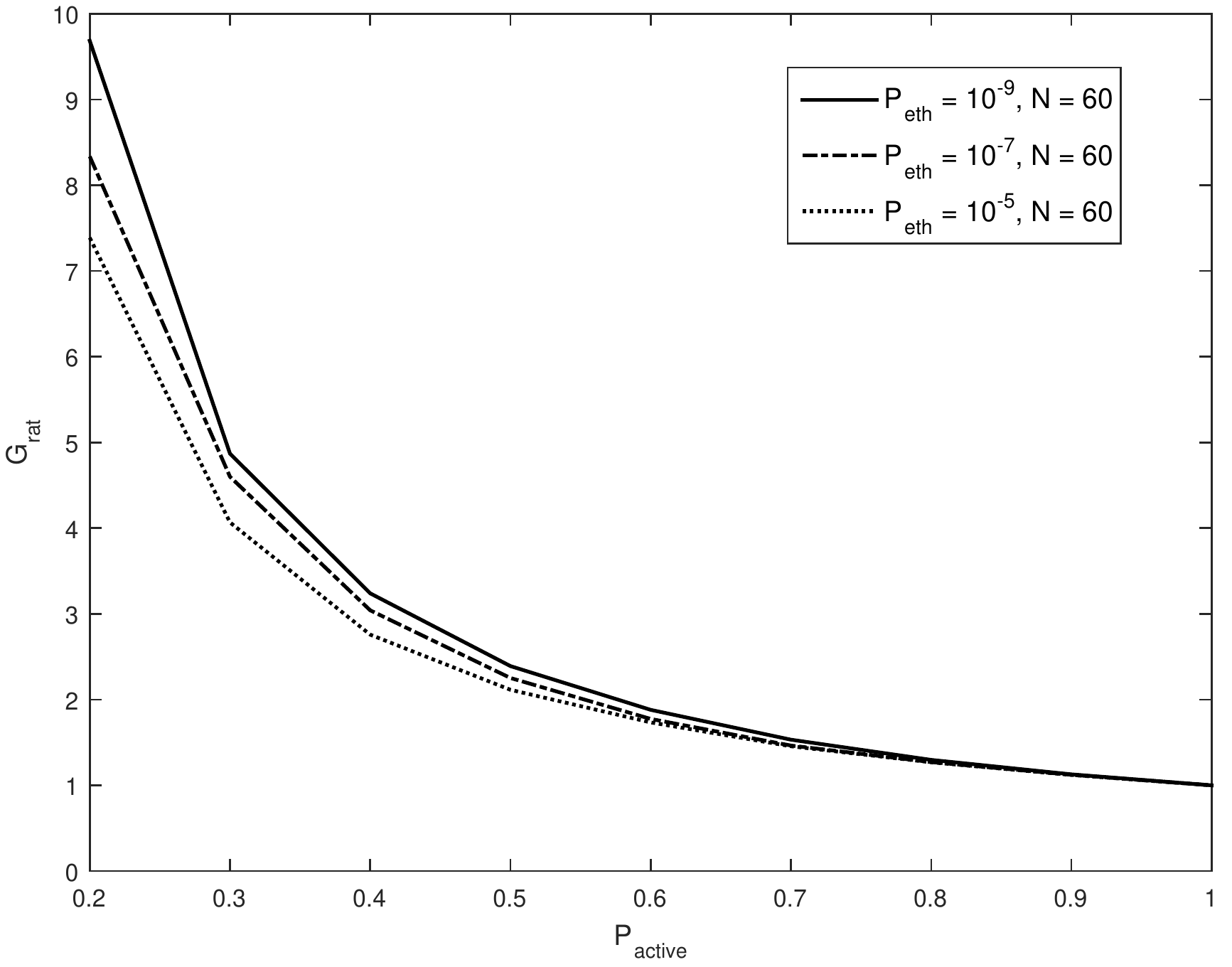}}
\center{\caption{\label{rate} Transmission rate gain $G_{rat}$ versus active probability $P_{active}$ for various values of BER threshold $P_{e_{th}}$ and maximum number of users $N=60$. $G_{rat}$ is defined as the ratio of the transmission rate in an adaptive OCDMA-PON system running rate-efficient code allocation Alg. \ref{flowchart2} to the transmission rate in a simple OCDMA-PON system that has no code reallocation.}}
\end{figure}

To evaluate the performance of the rate-efficient code allocation algorithm, assume that each user is active in code reallocation interval $T$ with the probability of $P_{active}$. Also assumed is that $P_{active}$'s are i.i.d over different code reallocation intervals and different users. Fig. \ref{rate} shows transmission rate gain $G_{rat}$ versus $P_{active}$ for different values of error threshold $P_{e_{th}}$ in a 1D OCDMA-PON that has at most $N=60$ active users. $G_{rat}$ equals to the ratio of the transmission rate in the proposed adative OCDMA-PON system running Alg. \ref{flowchart2} to the transmisson rate in the simple OCDMA system that has no code reallocation. Running Alg. \ref{flowchart2}, the transmission rate increases for lower values of $P_{active}$ since the control unit shares the unused resources of the inactive users among active ones to increase transmission rate. For instance, when $P_{e_{th}} = 10^{-7}$ and $P_{active} = 0.5$, $G_{rat} = 2.25$ which means that the transmission rate is $2.25$ times higher than the scenario in which no adaptive code allocation is available. Furthermore, the amount of transmission rate improvement is enhanced for lower values of $P_{e_{th}}$.

\begin{figure}
\center{\includegraphics[scale=0.47]{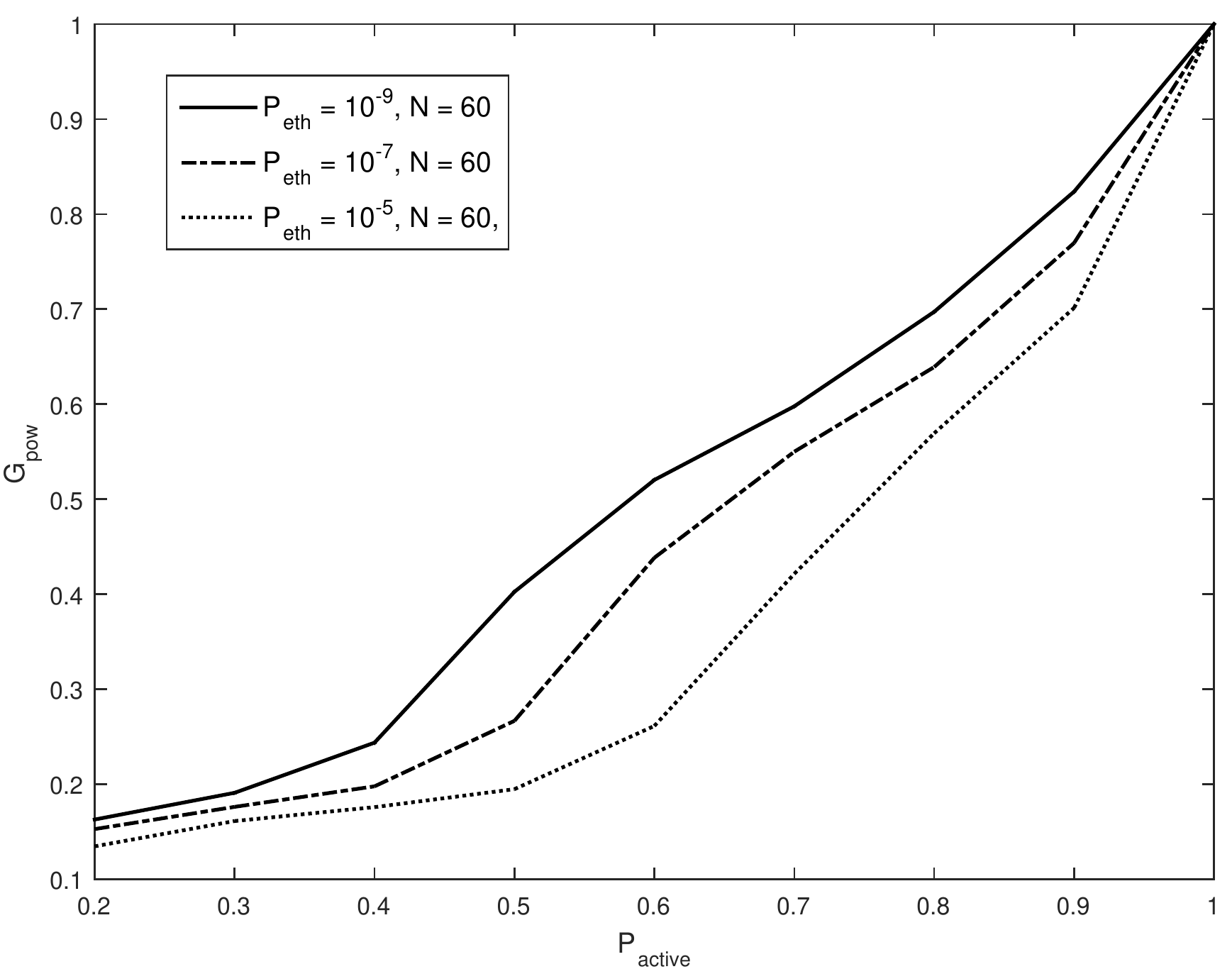}}
\center{\caption{\label{power}optical power consumption gain $G_{pow}$ versus active probability $P_{active}$ for various values of BER threshold $P_{e_{th}}$ and maximum number of users $N=60$. $G_{pow}$ is defined as the ratio of the average optical power consumption in an adaptive OCDMA-PON system running power-efficient code allocation Alg. \ref{flowchart3} to the average optical power consumption in a simple OCDMA-PON system that has no code reallocation.}}
\end{figure}

Fig. \ref{power} shows optical power consumption gain $G_{pow}$ versus $P_{active}$ for various values of error threshold $P_{e_{th}}$ in the adaptive 1D OCDMA-PON system illustrated in Fig. \ref{sdn_fig} that has at most $N=60$ users and runs power-efficient code allocation Alg. \ref{flowchart3}. We define $G_{pow}$ as the ratio of the average optical power consumption in the proposed adative OCDMA-PON system running Alg. \ref{flowchart3} to the average optical power consumption in the simple OCDMA-PON system that has no code reallocation. Clearly, the amount of optical power consumption reduces as $P_{active}$ decreases and the amount of power saving improves if the system can tolerate higher BER values. For instance, when $P_{e_{th}} = 10^{-7}$ and $P_{active} = 0.5$, the optical power consumption is reduced by a factor of $G_{pow}=0.27$ for adaptive code reallocation scenario compared to fixed code assignment one. Regarding the lightness of the code allocation algorithms, control unit can dynamically adapts resources and codes to activity of the users and dynamic fluctuation of traffics. Note that it is a straight forward way to combine the ideas behind our green code allocation and available green resource assignment methods (like one that puts network elements in sleep modes) to more reduce power consumption.
\section{Conclusion}\label{sec_VII}
In this paper, we provide an approximate simple expression for BER of a multi-class OCDMA system and verify its accuracy against the previously published complex BER expressions. The proposed approximation is valid for 1D/2D OCDMA systems with generalized or strict optical orthogonal codes. It also supports multi-class OCDMA systems having QoS classes differentiated by system level techniques such as multi-length variable-weight, multi-level and diversity signaling. We then provide some examples to show the applications of the derived BER equation in providing guidelines for analysis and design of OCDMA-based systems. We especially formulate two design problems for maximizing transmission rate or minimizing power consumption under constraints of code design, code cardinality and targeted system performance. We also use the BER equation to interpret  the sensitivity of BER performance to various system parameters and use the results to propose two adaptive code allocation algorithms for OCDMA-PONs. Simulation results show that regarding the amount of traffic fluctuations, randomness of the users activities and system BER, the proposed algorithms can increase transmission rate by a factor of $1-10$ or decrease power consumption by a factor of $0.1-1$. The proposed BER expression can be used in analysis and design of various types of  OCDMA systems and there is a good potential for its application in many other scenarios.

%
\appendices
\section{BER Approximation}\label{appen}
\begin{figure*}[!t]
\normalsize
\setcounter{MYtempeqncnt}{\value{equation}}
\setcounter{equation}{14}
\begin{align}\label{appen_7}
P_e  \leqslant  \frac{1}{2} \sum\limits_{i=0}^{W} (-1)^i {W \choose i}\sum\limits_{k_0+...+k_\lambda = N-1} {N-1 \choose k_0, k_1, ..., k_\lambda} \prod\limits_{t=0}^{\lambda} (f_t i^t)^{k_t} = \frac{1}{2} \sum\limits_{k_0+k_1+...+k_\lambda = N-1} {N-1 \choose k_0, k_1, ..., k_\lambda}\prod\limits_{t=1}^{\lambda} \vert f_t \vert ^{k_t} \sum\limits_{i=0}^{W} (-1)^i {W \choose i} (-i) ^ {\sum\limits_{t=1}^{\lambda} tk_t}
\end{align}
\setcounter{equation}{\value{MYtempeqncnt}}
\hrulefill
\vspace*{4pt}
\end{figure*}
\begin{figure*}[!t]
\normalsize
\setcounter{MYtempeqncnt}{\value{equation}}
\setcounter{equation}{23}
\begin{align}\label{appen_17}
& P_{e(1)} \leqslant  \frac{1}{2} \sum\limits_{i=0}^{W_1} (-1)^i {W_1 \choose i} \sum\limits_{m_0+...+m_{\lambda_{11}} = N_1-1} {N_1-1 \choose m_0,  ..., m_{\lambda_{11}}}\prod\limits_{s=0}^{\lambda_{11}} (f_s i^s)^{m_s} \sum\limits_{n_0+...+n_{\lambda_{12}} = N_2} {N_2 \choose n_0, ..., n_{\lambda_{12}}} \prod\limits_{t=0}^{\lambda_{12}} (g_t i^t)^{n_t} \\
\nonumber & = \frac{1}{2} \sum\limits_{m_0+...+m_{\lambda_{11}} = N_1-1} {N_1-1 \choose m_0, ..., m_{\lambda_{11}}} \prod\limits_{s=0}^{\lambda_{11}} \vert f_s \vert ^{m_s} \sum\limits_{n_0+...+n_{\lambda_{12}} = N_2} {N_2 \choose n_0, ..., n_{\lambda_{12}}} \prod\limits_{t=0}^{\lambda_{12}} \vert g_t \vert ^{n_t} \sum\limits_{i=0}^{W_1} (-1)^i {W_1 \choose i} (-i) ^ {\sum\limits_{s=1}^{\lambda_{11}} sm_s+\sum\limits_{t=1}^{\lambda_{12}} tn_t}
\end{align}
\hrulefill
\begin{align}\label{appen_18}
\nonumber P_{e(1)} &\approx \frac{W_1!}{2} \sum\limits_{u+v=W} \sum\limits_{m_1+2m_2+\cdots+\lambda_{11}m_{\lambda_{11}}=u}\frac{1}{m_1!\cdots m_{\lambda_{11}}!}\prod\limits_{s=0}^{\lambda_{11}} \vert N_1f_s \vert ^{m_s}\sum\limits_{n_1+2n_2+\cdots+\lambda_{12}n_{\lambda_{12}}=v}\frac{1}{n_1!\cdots n_{\lambda_{12}}!}\prod\limits_{t=0}^{\lambda_{12}} \vert N_2g_t \vert ^{n_t}\\
 & \approx \frac{1}{2}\sum\limits_{u+v=W} \frac{W_1!}{u!v!}\vert N_1f_{\lambda_{11}}\vert ^ {\frac{u}{\lambda_{11}}}\vert N_2g_{\lambda_{12}}\vert ^ {\frac{v}{\lambda_{12}}} =\frac{1}{2} (\vert N_1f_{\lambda_{11}}\vert ^ {\frac{1}{\lambda_{11}}}+\vert N_2g_{\lambda_{12}}\vert ^ {\frac{1}{\lambda_{12}}})^{W_1}
\end{align}
\hrulefill
\begin{align}\label{appen_19}
P_{e(1)} & \approx \frac{1}{2}\bigg[\Big(\frac{N_1W_1^2}{2M\lambda_{11} L_1W_1\cdots(W_1-\lambda_{11}+1)}\Big)^{\frac{1}{\lambda_{11}}}+\Big(\frac{N_2W_1W_2}{2M\lambda_{12} L_2W_1\cdots(W_1-\lambda_{12}+1)}\Big)^{\frac{1}{\lambda_{12}}}\bigg]^{W_1}\lesssim \frac{1}{2}\bigg[\Big(\frac{N_1W_1}{2M\lambda_{11} L_1}\Big)^{\frac{1}{\lambda_{11}}}+\Big(\frac{N_2W_2}{2M\lambda_{12} L_2}\Big)^{\frac{1}{\lambda_{12}}}\bigg]^{W_1}
\end{align}
\setcounter{equation}{\value{MYtempeqncnt}}
\hrulefill
\vspace*{4pt}
\end{figure*}
Through this appendix, we derive an approximated expression for BER of a $K$-class OCDMA system characterized by $\text{OCDMA}(K, M, \bold{N}, \bold{L}, \bold{W}, \bold{\Gamma}, \bold{C}, \bold{B})$. We first derive the approximation for a simple $1$-class OCDMA system $\text{OCDMA}(1, N, L, W, \lambda, 1, 1)$ and then extend it to work for a general $2$-class and $K$-class OCDMA system. The following identity plays a key role in deriving the approximations \cite{Div}:
\begin{equation}\label{appen_1}
\sum\limits_{i=0}^{W}(-1)^i {W \choose i} (-i)^t = 
\begin{cases}
0 &\mbox{if } \hspace{2mm} t = 0, \cdots, W-1 \\
W! & \mbox{if } \hspace{2mm} t = W 
\end{cases}
\end{equation}
\subsection{$1$-class OCDMA system}
The BER of a $1$-class 2D OCDMA system is upper limited by \cite{ref_9} (for 1D OCDMA system, simply set $M = 1$):
\begin{equation}\label{appen_3}
P_e  \leqslant \frac{1}{2}\sum\limits_{i=0}^{W} (-1)^i {W \choose i} [f(i)]^{N-1}
\end{equation}
where:
\begin{equation}\label{appen_4}
f(i) = 1- \frac{W^2}{2M\lambda L}\Big[1-\frac{{W-i \choose \lambda}}{{W \choose \lambda}}\Big]
\end{equation}
Obviously, $f(i)$ can be considered as a $\lambda$-th order polynomial with coefficients $f_t, t = 0, \cdots, \lambda$  and $\lambda$ is usually chosen between $1$ to $3$ \cite{ref_9}. The constant coefficient of $f(i)$ is $f_0 = 1$. In practical implementations of an OCDMA system, the code length $L$ is usually a large value and therefore, $\vert f_t \vert \ll 1, t=1,2,\cdots, \lambda$. Furthermore, one can check that $f_t$ is positive and negative for even and odd values of $t$, i.e. $f_t = (-1)^t \vert f_t \vert$. Now, we use Newton multinomial expansion to reform \eqref{appen_3}, \addtocounter{equation}{1}
as can be seen in (\ref{appen_7}) at the top of the next page.

Noting the identity in \eqref{appen_1}, the first non-zero term of \eqref{appen_7} is obtained when $\sum\limits_{t=1}^{\lambda} tk_t=W$. We use this first non-zero term to approximate BER:
\begin{align}\label{appen_8}
P_e \approx \frac{W!}{2} \sum\limits_{\substack{k_0+k_1+...+k_\lambda = N-1 \\ k_1+2k_2+...+\lambda k_\lambda = W}} {N-1 \choose k_0, k_1, ..., k_\lambda} \prod\limits_{t=1}^{\lambda} \vert f_t \vert ^{k_t} 
\end{align}
For sufficiently large values of $N$, we approximately have ${N-1 \choose k_0, k_1, ..., k_\lambda} \approx \frac{N^{k_1+\cdots+k_\lambda}}{k_1!...k_\lambda !}$. Therefore:
\begin{align}\label{appen_9}
P_e \approx \frac{1}{2} \sum\limits_{ k_1+2k_2+...+\lambda k_\lambda = W}\frac{W!}{k_1!...k_\lambda !} \prod\limits_{t=1}^{\lambda} \vert Nf_t \vert ^{k_t} 
\end{align}
Defining $u_i = i k_i, i = 1,2, \cdots, \lambda$ and noting that $\vert f_t \vert \ll 1$, we have:
\begin{align}\label{appen_10}
\nonumber P_e & \approx \frac{1}{2}\sum\limits_{u_1+...+u_\lambda = W} \frac{W!}{(u_1)!...(\frac{u_\lambda }{\lambda}!)} \prod\limits_{t=1}^{\lambda} \vert Nf_t \vert ^{\frac{u_t}{t}} \\
& \approx \frac{1}{2}\Big(\sum\limits_{t=1}^{\lambda} \vert Nf_t \vert ^{\frac{1}{t}}\Big)^W \approx \frac{1}{2}\vert Nf_\lambda\vert^{\frac{W}{\lambda}}
\end{align}
Substituting the value of $f_\lambda$ in \eqref{appen_10} results in:
\begin{align}\label{appen_11}
P_e \approx \frac{1}{2}\Big(\frac{NW^2}{2M\lambda LW\cdots(W-\lambda+1)}\Big)^{\frac{W}{\lambda}} \lesssim \frac{1}{2}\Big(\frac{NW}{2M\lambda L}\Big)^{\frac{W}{\lambda}} 
\end{align}
\subsection{$2$-class OCDMA system}
Now, we derive the approximation for BER of the first class in an $\text{OCDMA}(2, M, \bold{N}, \bold{L}, \bold{W}, \bold{\Gamma}, \bold{1}, \bold{1})$ system. The same method can be applied to the second class to yield its corresponding BER approximation. BER of the first class of the desired $2$-class OCDMA system is upper limited by \cite{MS}:
\begin{equation}\label{appen_12}
P_{e(1)}  \leqslant \frac{1}{2}\sum\limits_{i=0}^{W_1} (-1)^i {W_1 \choose i} [f(i)]^{N_1-1}[g(i)]^{N_2}
\end{equation}
where:
\begin{equation}\label{appen_13}
f(i) = 1+ \frac{W_1^2}{2M\lambda_{11} L_1{W_1 \choose \lambda_{11}}}\sum\limits_{j=1}^{\lambda_{11}}(-1)^j {i \choose j}{W_1-\lambda_{11} \choose \lambda_{11}-j}
\end{equation}
\begin{equation}\label{appen_14}
g(i) = 1+ \frac{W_1W_2}{2M\lambda_{12} L_2 {W_1 \choose \lambda_{12}}}\sum\limits_{j=1}^{\lambda_{12}}(-1)^j {i \choose j}{W_1-\lambda_{12} \choose \lambda_{12}-j}
\end{equation}
Again, $f(i)$  and $g(i)$ can be considered as two polynomials having orders of $\lambda_{11}$ and $\lambda_{12}$ with coefficients $f_s, s =0, \cdots, \lambda_{11}$ and $g_t, t=0, \cdots, \lambda_{12}$, respectively. Similar to the previous derivation, we have $f_0 =1$, $g_0 = 1$, $f_s = (-1)^s \vert f_s \vert$ and $g_t = (-1)^t \vert g_t \vert$.
As shown in \eqref{appen_17} at the top of the page, we use Newton multinomial expansion to reform \eqref{appen_12}.
Assume $u = \sum\limits_{s=1}^{\lambda_{11}} sm_s$ and $v=\sum\limits_{t=1}^{\lambda_{12}} tn_t$. Noting the identity of \eqref{appen_1}, the first non-zero term of \eqref{appen_17} is obtained when $u+v=W$. We proceed with the same way as the previous sub-section to approximate BER with its first non-zero term, as can be seen in \eqref{appen_18} at the top of the page.
Substituting the values of $f_{\lambda_{11}}$ and $g_{\lambda_{12}}$ in \eqref{appen_18} results in the desired approximation expression, shown in \eqref{appen_19} at the top of the page.

\subsection{$K$-class OCDMA System}
The same procedure as previous sub-sections can be used to provide the following approximated expression for BER of an $\text{OCDMA}(K, M, \bold{N}, \bold{L}, \bold{W}, \bold{\Gamma}, \bold{C}, \bold{B})$ system:
\addtocounter{equation}{3}
\begin{align}\label{appen_19}
P_{e(k)}\lesssim \frac{1}{2}\bigg[\sum\limits_{i=1}^{K}\Big(\frac{N_iW_i}{2M\lambda_{ki} L_i}\Big)^{\frac{c_{ki}}{\lambda_{ki}}}\bigg]^{W_kb_k}
\end{align}
\ifCLASSOPTIONcaptionsoff
  \newpage
\fi

\end{document}